\documentclass[useAMS,usenatbib,onecolumn]{mn2e}
\usepackage{graphicx}
\usepackage{amsfonts}
\title[On-off intermittency and amplitude-phase synchronization in Keplerian shear flows]{On-off intermittency and amplitude-phase synchronization in Keplerian shear flows}
\author[R. A. Miranda, E. L. Rempel and A. C.-L. Chian]{R. A. Miranda$^{1,2}$\thanks{E-mail: rmiracer@unb.br},
 E. L. Rempel$^{2,3}$ and A. C.-L. Chian$^{3,4}$\\
$^{1}$University of Bras\'ilia (UnB), Faculty UnB-Gama, and Plasma Physics Laboratory, Institute of Physics, \\
 Bras\'ilia, DF 70910-900, Brazil.\\
$^{2}$Institute of Aeronautical Technology (ITA), S\~ao Jos\'e dos Campos, SP 12228-900, Brazil.\\
$^{3}$National Institute for Space Research (INPE), P. O. Box 515, S\~ao Jos\'e dos Campos, SP 12227-010, Brazil.\\
$^{4}$School of Mathematical Sciences, University of Adelaide, Adelaide, SA 5005, Australia.
}
\begin{document}


\pagerange{\pageref{firstpage}--\pageref{lastpage}} \pubyear{2002}

\maketitle

\label{firstpage}

\begin{abstract}
We study the development of coherent structures in local simulations of the magnetorotational instability in accretion discs in regimes of on-off intermittency. In a previous paper [Chian et al., Phys. Rev. Lett. 104, 254102 (2010)], we have shown that the laminar and bursty states due to the on-off spatiotemporal intermittency in a one-dimensional model of nonlinear waves correspond, respectively, to nonattracting coherent structures with higher and lower degrees of amplitude-phase synchronization. In this paper we extend these results to a three-dimensional model of magnetized Keplerian shear flows. Keeping the kinetic Reynolds number and the magnetic Prandtl number fixed, we investigate two different intermittent regimes by varying the plasma beta parameter. The first regime is characterized by turbulent patterns interrupted by the recurrent emergence of a large-scale coherent structure known as two-channel flow, where the state of the system can be described by a single Fourier mode. The second regime is dominated by the turbulence with sporadic emergence of coherent structures with shapes that are reminiscent of a perturbed channel flow.  By computing the Fourier power and phase spectral entropies in three-dimensions, we show that the large-scale coherent structures are characterized by a high degree of amplitude-phase synchronization.
\end{abstract}


\section{Introduction}

Accretion discs are disc-like structures formed by the accumulation of mass onto a central object, such as a young star, due to gravitational attraction \citep{biskamp:2003}. The infalling material tends to rotate about the center of gravity due to conservation of angular momentum. In a Keplerian disc, the angular momentum decreases with decreasing radius, hence a transport mechanism must be provided to remove the excess of angular momentum due to the infalling material. The most promising mechanism is the generation of turbulence in the presence of a background magnetic field through the magnetorotational instability (MRI) \citep{balbus_hawley:1991}.

Theoretical studies of turbulence in plasmas have benefited from numerical simulations. However, the large number of active scales in a turbulent flow requires simulations with high resolution in space and time, which is computationally expensive. In the case of turbulent accretion discs, it is convenient to use a local model called the shearing box formalism \citep{hawley_balbus:1992,hawley_etal:1995}. In this model the simulation domain is restricted to a box centered at a fiducial disc radius of dimensions much smaller than the size of the disc, allowing to focus on local phenomena such as the growth of unstable modes via MRI, turbulent patterns and energy cascades without requiring an extremely high computational effort.

Since the numerical work by \citet{hawley_balbus:1992}, numerical simulations of magnetized Keplerian shear flows using the shearing box formalism have shown the formation of large-scale coherent structures called two-channel flows. These flows arise due to the MRI acting on modes with low wavenumbers under certain regimes \citep{hawley_balbus:1992, sano_inutsuka:2001, lesur_longaretti:2007}, and represent exact solutions of the incompressible magnetohydrodynamic equations \citep{goodman_xu:1994}. The unstable modes grow exponentially and then break, transferring energy to the smaller scales. Different mechanisms have been proposed to explain the destruction of two-channel flows, including parasitic instabilities \citep{goodman_xu:1994, pessah_goodman:2009} and low-level turbulence \citep{latter_etal:2009}. The generation of two-channel flows via MRI makes the shearing box model a test-case to study coherent structures in turbulent flows. Coherent structures are responsible for several important features observed in space and astrophysical plasmas such as non-Gaussianity \citep{sorrisovalvo_etal:2001, chian_miranda:2009,hilyblant_falgarone:2009,falcetagoncalves_etal:2014}, multifractality \citep{bershadskii_sreenivasan:2004, bruno_etal:2007, miranda_etal:2013} and phase synchronization among scales \citep{koga_etal:2007, chian_miranda:2009}. 

In this paper we analyze three-dimensional (3D) numerical simulations of MRI-driven turbulence in magnetized Keplerian shear flows using a theoretical framework based on the information entropy theory developed by \citet{shannon:1949}, applied to the amplitude and phase information of Fourier modes in the 3D spectral space. The power spectral entropy, which measures the degree of disorder of the Fourier amplitudes, has previously been applied to study the onset of spatiotemporal intermittency in nonlinear models of drift waves in plasmas \citep{rempel_etal:2007, rempel_etal:2009a} and in a nonlinear 3D model of a mean-field dynamo \citep{rempel_etal:2009b,rempel_etal:2011}. The phase spectral entropy, which measures the degree of disorder of phase differences, has been applied in a nonlinear model of coupled oscillators \citep{lai_etal:2006} and in a model of cosmological density perturbations \citep{chiang_coles:2000}. \cite{chian_etal:2010} demonstrated the duality of the amplitude and phase synchronization in on-off spatiotemporal intermittency by applying the power and phase spectral entropies to a one-dimensional model of nonlinear long-waves. They found that, in the spatiotemporally intermittent regime, the degree of amplitude-phase synchronization is higher during the ``on'' states corresponding to coherent structures which are regular in space and chaotic in time, whereas the degree of amplitude-phase synchronization is lower during the ``off'' states associated with coherent structures which are irregular in space and time. Here, we seek to extend this framework to characterize the recurrent formation of channel-flow coherent structures in on-off intermittency in 3D simulations of magnetized Keplerian shear flows.

This paper is organized as follows. In Section \ref{sec_eqs} we review the equations which describe the dynamics of a magnetized Keplerian disc in the shearing box formalism. In Section \ref{sec_entropies} we describe the implementation details of the Fourier power and phase spectral entropies taking into account the shearing boundary conditions. We proceed with the numerical analysis in Section \ref{sec_numerical}, followed by a discussion in Section \ref{sec_discussion}. Our conclusion is presented in Section \ref{sec_conclusion}.

\section[]{The shearing-box equations} \label{sec_eqs}
The dynamics of a box moving with the angular velocity of a magnetized Keplerian disc can be represented in Cartesian coordinates as follows

  \begin{eqnarray}
  \frac{\partial \mathbf{u}}{\partial t} + (\mathbf{u} \cdot \nabla) \mathbf{u} & = & -\frac{1}{\rho} \nabla P + \frac{1}{\mu_0 \rho} (\nabla \times \mathbf{B}) \times \mathbf{B} - 2 \mathbf{\Omega} \times \mathbf{u} + 2\Omega S x \mathbf{\hat{x}} + \nu \nabla^2 \mathbf{u} \label{eq_mhd1}\\
  \frac{d\mathbf{B}}{dt} & = & \nabla \times (\mathbf{u} \times \mathbf{B}) + \eta \nabla^2 \mathbf{B} \label{eq_mhd2}\\
  \nabla \cdot \mathbf{u} & = & 0 \\
  \nabla \cdot \mathbf{B} & = & 0 \label{eq_mhd4}
  \end{eqnarray}

\noindent where $\rho$ represents the gas density, $P$ represents the pressure, $\mu_0$ represents the magnetic permeability, $\Omega = r^{-3/2}$, $S = -r \partial \Omega/\partial r$, $\nu$ represents the kinematic viscosity and $\eta$ represents the magnetic diffusivity. The $x$ coordinate represents the radial direction, the $y$ coordinate is the azimuthal direction, and the $z$ coordinate represents the vertical direction. The fluid velocity can be decomposed as

\begin{equation} \label{eq_perturbation}
\mathbf{u} = \mathbf{u}_0 + \mathbf{V}
\end{equation}

\noindent where $\mathbf{u}_0 = - S x \mathbf{\hat{y}} $ represents the steady-state solution, $\mathbf{V}$ is the perturbation velocity field, and $S = (3/2)\Omega$ for a Keplerian flow. Inserting (\ref{eq_perturbation}) into Eqs. (\ref{eq_mhd1})-(\ref{eq_mhd2}) gives \citep{lesur_longaretti:2011}

  \begin{eqnarray}
  \frac{\partial \mathbf{V}}{\partial t} + (\mathbf{V} \cdot \nabla) \mathbf{V} & = & -\frac{1}{\rho}\nabla \psi + \frac{1}{\mu_0 \rho} \left( \mathbf{B} \cdot \nabla \right) \mathbf{B} + Sx \frac{\partial \mathbf{V}}{\partial y} + (S - 2\Omega) V_x \mathbf{\hat{y}} + 2 \Omega V_y \mathbf{\hat{x}} \nonumber \\
& &  + \nu \nabla^2 \mathbf{V} \label{eq_1}\\
  \frac{\partial \mathbf{B}}{\partial t} + (\mathbf{V} \cdot \nabla) \mathbf{B} & = & (\mathbf{B} \cdot \nabla) \mathbf{V} + Sx\frac{\partial \mathbf{B}}{\partial y} - S B_x \mathbf{\hat{y}} + \eta \nabla^2 \mathbf{B} \label{eq_4}
  \end{eqnarray}

\noindent where $\psi = (P + B^2/\mu_0)/\rho$ is the perturbation in the total pressure. Periodic boundary conditions can be enforced by transforming Eqs. (\ref{eq_1}) and (\ref{eq_4}) into a shearing frame $(X, Y, Z)$ described by \citep{umurhan_regev:2004,lesur_longaretti:2005}

  \begin{equation} \label{eq_shearingcoords}
    X = x, \qquad Y = y + Sx(t - t_0), \qquad Z = z, \qquad T = t - t_0,
  \end{equation}

\noindent thus $x$, $y$, $z$ represent the non-shearing coordinates. In this frame Eqs. (\ref{eq_1}) and (\ref{eq_4}) become \citep{lesur_longaretti:2011}

  \begin{eqnarray}
  \frac{\partial \mathbf{V}}{\partial t} + (\mathbf{V} \cdot \tilde{\nabla}) \mathbf{V} & = & -\frac{1}{\rho}\tilde{\nabla} \psi + \frac{1}{\mu_0 \rho} \left( \mathbf{B} \cdot \tilde{\nabla} \right) \mathbf{B} + (S - 2\Omega) V_x \mathbf{\hat{y}} + 2 \Omega V_y \mathbf{\hat{x}} \nonumber \\
& &  + \nu \tilde{\nabla}^2 \mathbf{V} \label{eq_nonshear1}\\
  \frac{\partial \mathbf{B}}{\partial t} + (\mathbf{V} \cdot \tilde{\nabla}) \mathbf{B} & = & (\mathbf{B} \cdot \tilde{\nabla}) \mathbf{V} - S B_x \mathbf{\hat{y}} + \eta \tilde{\nabla}^2 \mathbf{B}, \label{eq_nonshear2}
  \end{eqnarray}

\noindent where

  \begin{displaymath}
    \tilde{\nabla} = \left(\frac{\partial}{\partial X} + St \frac{\partial}{\partial Y}\right)\mathbf{\hat{x}} + \frac{\partial}{\partial Y}\mathbf{\hat{y}} + \frac{\partial}{\partial Z} \mathbf{\hat{z}}.
  \end{displaymath}

\noindent The explicit spatial dependence of Eqs. (\ref{eq_1}) and (\ref{eq_4}) due to the terms proportional to $x$ is thus removed and the physical quantities in Eqs. (\ref{eq_nonshear1}) and (\ref{eq_nonshear2}) can be decomposed into Fourier modes to enforce the periodic boundary conditions. 

The kinetic Reynolds number $Re$ and the magnetic Reynolds number $Rm$ can be defined as follows

  \begin{eqnarray*}
    Re & = & \frac{S d^2}{\nu}\\
    Rm & = & \frac{S d^2}{\eta}
  \end{eqnarray*}

\noindent where $d$ represents the shear-wise size of the box. The plasma beta parameter can be defined as \citep{lesur_longaretti:2007}

\begin{equation}
  \beta = \frac{S^2 d^2}{V_A^2}
\end{equation}

\noindent where $V_A^2 = B_0^2/(\mu_0 \rho)$ represents the square of the Alfv\'en speed, and $B_0$ is the background magnetic field.

We solve Eqs. (\ref{eq_nonshear1})-(\ref{eq_nonshear2}) using the pseudospectral code described by \citet{lesur_longaretti:2007, lesur_longaretti:2011} with a 2/3 dealiasing rule. The initial conditions are set to random white noise with an amplitude of 0.1 on all scales. We set the box $L_x$:$L_y$:$L_z$ aspect ratio to 1:4:1 with a resolution $N_x \times N_y\times N_z = 64 \times 128 \times 64$, $\rho = \mu_0 = S$ = 1, $\Omega$ = 2/3, $Re$ = 3200 and the magnetic Prandtl number $Pm = Re/Rm$ = 1. The background magnetic field is parallel to the $z$ axis. Note that this choice of parameters allows us to obtain the magnitude of the background magnetic field as $B_0 = 1/\sqrt{\beta}$. We focus on two values of the plasma $\beta$ parameter, namely, $\beta = 30$ ($B_0 \sim 0.18157$) and $\beta = 100$ ($B_0 = 0.1$) which are described in detail in Section \ref{sec_numerical}.

\section{The power-phase spectral entropy} \label{sec_entropies}

The Fourier decomposition of a three-dimensional scalar field $\theta(\mathbf{x}, t)$ can be written as

  \begin{equation} \label{eq_scalarfieldinfourier}
    \theta(\mathbf{x}, t) = \sum_{\mathbf{k}} \hat{\theta}_\mathbf{k}(t) \exp(i \mathbf{k}\cdot \mathbf{x})
  \end{equation}

\noindent where $\mathbf{x} = (x, y, z)$, $\mathbf{k} = (k_x, k_y, k_z)$, $k_i = (2\pi/L_i)n_i$, $n_i = 1, 2, ..., N_i$ ($i = x, y, z$), the symbol $\sum_{\mathbf{k}}$ denotes the triple sum $\sum_{k_x} \sum_{k_y} \sum_{k_z}$, and $\hat{\theta} (t)$ represents the complex Fourier coefficient. For each Fourier coefficient, we define its amplitude

  \begin{equation}
    |\hat{\theta}_\mathbf{k}(t)| = \sqrt{ \hat{\theta}_\mathbf{k}(t)\hat{\theta}_\mathbf{k}^*(t)}
  \end{equation}

\noindent and phase

  \begin{equation}
    \phi_\mathbf{k}(t) = \arctan \left( \frac{Im\{ \hat{\theta}_\mathbf{k}(t)\}}{Re\{\hat{\theta}_\mathbf{k}(t)\}} \right)
  \end{equation}

\noindent We quantify the degree of amplitude synchronization due to multiscale interactions by the Fourier power spectral entropy

  \begin{equation} \label{eq_ska}
    S_k^A(t) = - \sum_{\mathbf{k}} p\left( \hat{\theta}_{\mathbf{k}}(t) \right) \ln \left[ p \left( \hat{\theta}_{\mathbf{k}} (t) \right) \right]
  \end{equation}

\noindent where

  \begin{equation}
    p\left(\hat{\theta}_\mathbf{k}(t)\right) = \frac{|\hat{\theta}_\mathbf{k} (t) |^2}{\sum_\mathbf{k} |\hat{\theta}_\mathbf{k} (t) |^2}
  \end{equation}

\noindent is the relative weight of Fourier mode $\mathbf{k}$. Note that $p(\hat{\theta}_\mathbf{k}(t)) \in [0, 1]$, and $\sum_\mathbf{k} p(\hat{\theta}_\mathbf{k}(t)) = 1$. If $p(\hat{\theta}_\mathbf{k}(t)) = 1$ for a given $\mathbf{k}$, then $S_k^A(t) = 0$, which represents a perfectly ordered state for the amplitudes. If the energy is uniformly distributed across Fourier modes, then $p(\hat{\theta}_\mathbf{k}(t)) = 1/(N_x N_y N_z) \hspace{2ex} \forall \mathbf{k}$, and $S_k^A(t) = \ln(N_x N_y N_z)$, which is the maximum value of the power spectral entropy.

We quantify the degree of phase synchronization related to multiscale interactions as follows \citep{chiang_coles:2000}. Let us represent two neighbouring (or adjacent) modes in the $k_i$ direction by $k_i = (2\pi/L_i)n_i$ and $k_{i + 1} = (2\pi/L_i)(n_i + 1)$ ($i = x, y, z$). We compute the Fourier phase differences between neighbouring (or adjacent) modes in the $k_x$ direction by

  \begin{equation} \label{eq_deltaphikx}
    \delta \phi_{k_x}(t) = \phi(k_{x + 1}, k_y, k_z) - \phi(k_x, k_y, k_z)
  \end{equation}

\noindent and the Fourier phase differences in the $k_z$ direction by

  \begin{equation} \label{eq_deltaphikz}
    \delta \phi_{k_z}(t) = \phi(k_x, k_y, k_{z + 1}) - \phi(k_x, k_y, k_z)
  \end{equation}

\noindent We avoid computing the phase differences in the $k_y$ direction due to the shearing boundary conditions. This is explained below. The Fourier phase spectral entropy in the $k_x$ direction is given by

  \begin{equation} \label{eq_sphikx}
    S_{k_x}^\phi(t) = - \sum_{n_x=1}^{N_{\delta \phi}} P\left( \delta \phi_{\left( 2\pi n_x/L_x \right)}(t) \right) \ln \left[ P \left( \delta \phi_{\left( 2\pi n_x/L_x \right)}(t) \right) \right]
  \end{equation}

\noindent where $P$ is the probability distribution function (PDF) of $\delta \phi_{k_x}(t) = \delta \phi_{\left( 2\pi n_x/L_x \right)}(t)$, $n_x = 1, 2, ..., N_{\delta \phi}$, and $N_{\delta \phi}$ is the total number of phase differences which depends on an imposed threshold explained below. The PDF can be implemented by dividing the interval $[-\pi, \pi]$ into subintervals or bins, and constructing a normalized histogram of $\delta \phi_{k_x}$ by $P(\delta \phi_{k_x}) = b_i/N_{\delta \phi}$, where $b_i$ is the number of times $\delta \phi_{k_x}$ falls into the $i$th bin. Hence, $P(\delta \phi_{k_x}) \in [0, 1]$ and $\sum_{n_x=1}^{N_{\delta \phi}} P(\delta \phi_{\left( 2\pi n_x/L_x \right)}) = 1$. If $P(\delta \phi_{k_x}) = 1$ for a given $\delta \phi_{k_x}$, then $S_k^\phi(t) = 0$ which indicates a perfect phase synchronization. If $P(\delta \phi_{k_x})$ is a uniform distribution, then $S_k^\phi = \ln(N_{\delta \phi})$, which represents the maximum value of the Fourier phase spectral entropy. Likewise, the Fourier phase spectral entropy in the $k_z$ direction is

  \begin{equation} \label{eq_sphikz}
    S_{k_z}^\phi(t) = - \sum_{n_z=1}^{N_{\delta \phi}} P\left( \delta \phi_{\left( 2\pi n_z/L_z \right)}(t) \right) \ln \left[ P \left( \delta \phi_{\left( 2\pi n_z/L_z \right)}(t) \right) \right]
  \end{equation}

\noindent Following \cite{chiang_coles:2000} we take the arithmetic mean to obtain

  \begin{equation} \label{eq_sphi}
    S_k^\phi(t) = \frac{S_{k_x}^\phi (t) + S_{k_z}^\phi (t)}{2}
  \end{equation}

\noindent which we call the Fourier phase spectral entropy. We restrict the sums in Eqs. (\ref{eq_sphikx}) and (\ref{eq_sphikz}) to Fourier modes with amplitude larger than a threshold value of $5 \times 10^{-3}$. This is due to the dealiasing procedure in which 2/3 of Fourier modes are set to zero during the computation of the nonlinear terms of Eqs. (\ref{eq_nonshear1})-(\ref{eq_nonshear2}). In addition, several modes have extremely small amplitudes during the formation of the two-channel flows at $\beta$ = 30. These modes can be regarded as numerical noise that may contribute to Eqs. (\ref{eq_sphikx}) and (\ref{eq_sphikz}) with random phases leading to an artificial increase of entropy of phase-differences.

The numerical implementation of the shearing sheet boundary conditions in Fourier space introduces some modifications with respect to a standard spectral method. Solving (\ref{eq_shearingcoords}) for the non-shearing coordinates one obtains

  \begin{equation} \label{eq_observerframe}
    x = X, \qquad y = Y - SXT, \qquad z = Z, \qquad t = T + t_0
  \end{equation}

\noindent Substituting (\ref{eq_observerframe}) into Eq. (\ref{eq_scalarfieldinfourier}) gives

  \begin{equation} 
    \theta(\mathbf{x}, t) = \sum_{\mathbf{k}} \hat{\theta}_\mathbf{k}(t) \exp[i(k_{\mathrm{eff}}X + k_y Y + k_z Z)]
  \end{equation}

\noindent where

  \begin{equation} \label{eq_keff}
    k_{\mathrm{eff}}(T) = k_x - S k_y T
  \end{equation}

\noindent is the time-dependent ``effective'' wavenumber in the $X$ direction. The Fourier wavevectors in the numerical implementation are given by

  \begin{equation} \label{eq_bigk}
    \mathbf{K} = (k_{\mathrm{eff}}, k_y, k_z)
  \end{equation}

\noindent For a fixed non-shearing wavevector $\mathbf{k} = (k_x, k_y, k_z)$, the modulus of $\mathbf{K}$ grows with $T$. This implies that the energy will shift towards negative $k_\mathrm{eff}$ for positive $k_y$, and towards positive $k_\mathrm{eff}$ for negative $k_y$. Since the numerical domain has a finite size, the shearing effect can lead to a loss of information. Note that, at $t = t_0$, the shearing and the non-shearing coordinates coincide. Therefore, one can avoid the loss of information by periodically transforming $\mathbf{K}$ to the non-shearing coordinates, resetting $T = 0$, and then using the resulting $\mathbf{k}$ wavevectors as initial conditions for $\mathbf{K}$ \citep{umurhan_regev:2004,lesur_longaretti:2005}. This remapping process is applied in every $T = L_y/(SL_x)$ time units \citep{lesur_longaretti:2009}.

\begin{figure}
  \includegraphics[width=\textwidth]{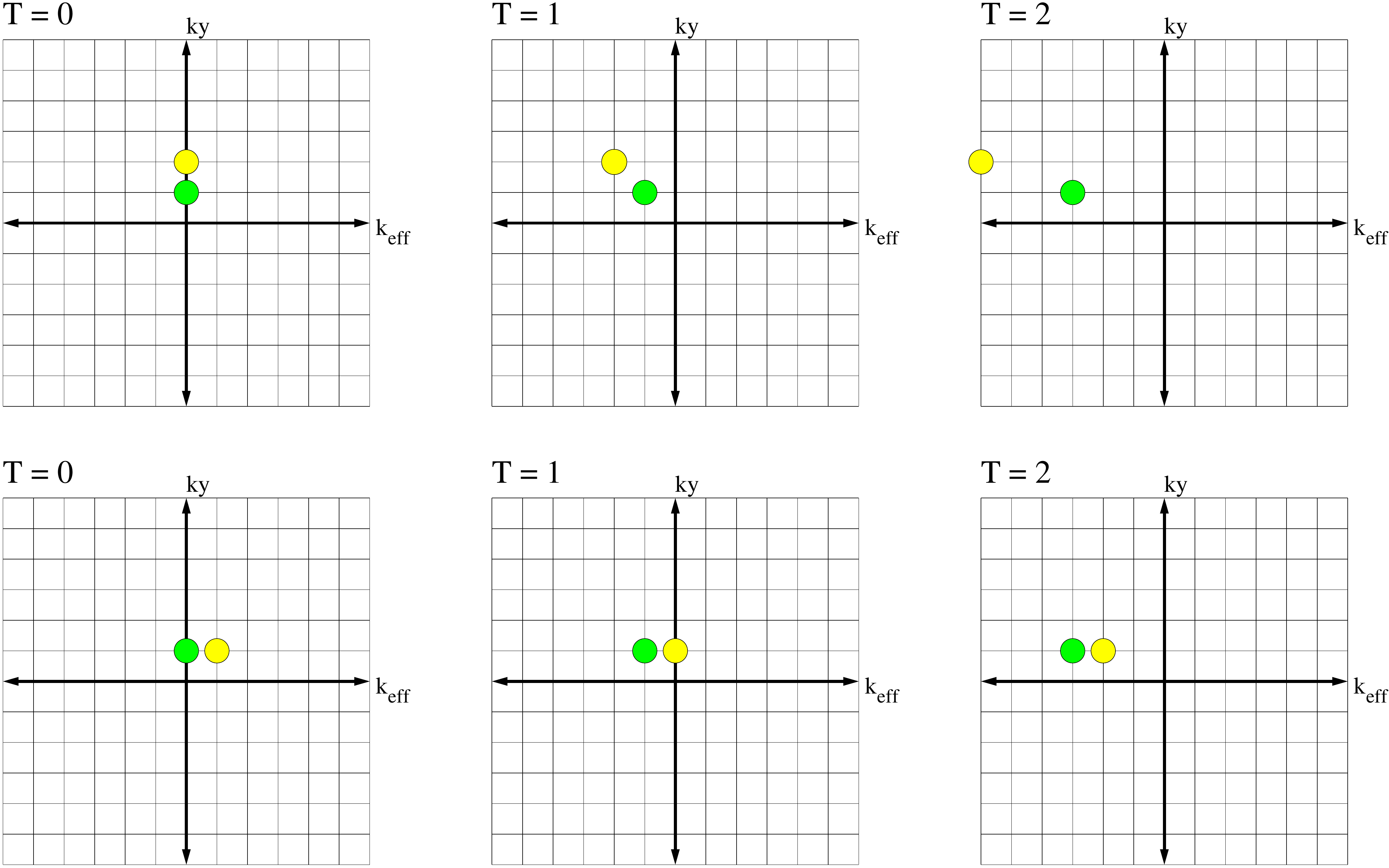}
  \caption{Schematic view of the plane $(k_{\mathrm{eff}}, k_y)$ at $k_z = 0$ in Fourier space, showing the effect of shearing boundary conditions on neighbouring (adjacent) wavevectors. Upper panels: wavevectors $\mathbf{K}_1$ = (0, 1, 0) (green) and $\mathbf{K}_2$ = (0, 2, 0) (yellow) at $T$ = 0, $T$ = 1 and $T$ = 2. Lower panels: wavevectors $\mathbf{K}_1$ = (0, 1, 0) (green) and $\mathbf{K}_2$ = (1, 1, 0) (yellow) at $T$ = 0, $T$ = 1 and $T$ = 2.}
  \label{fig2}
\end{figure}

Recall that we quantify phase synchronization using phase differences of adjacent modes (e.g., between $\phi(k_x, k_y, k_z)$ and $\phi(k_{x + 1}, k_y, k_z)$). The distance between two wavevectors with different values of the $k_y$ component will increase with time as an effect of the shearing sheet boundary conditions. For example, suppose that we compute the phase differences between wavevectors $\mathbf{k}_1 = (0, 1, 0)$ and $\mathbf{k}_2 = (0, 2, 0)$ which occupy adjacent positions in the nonshearing coordinates. Let $\mathbf{K}_1$ and $\mathbf{K}_2$ be their respective representation in the shearing coordinates. At $T = 0$, Eqs. (\ref{eq_keff}) and (\ref{eq_bigk}) give $\mathbf{K}_1 = \mathbf{k}_1$ and $\mathbf{K}_2 = \mathbf{k}_2$. The time evolution of $\mathbf{K}_1$ and $\mathbf{K}_2$ is illustrated in the upper panels of Fig. \ref{fig2}. At $T = 1$, Eqs. (\ref{eq_keff}) and (\ref{eq_bigk}) give $\mathbf{K}_1 = (-1, 1, 0)$ and $\mathbf{K}_2 = (-2, 2, 0)$.   At $T$ = 2, $\mathbf{K}_1 = (-3, 1, 0)$ and $\mathbf{K}_2 = (-6, 2, 0)$, which are non-contiguous. For this reason we avoid taking the phase differences in the $k_y$ direction.

We note that the distance between two wavevectors with the same value of the $k_y$ component does not increase, and as a result, the phase differences of neighbouring modes will not be affected in the $k_\mathrm{eff}$ and the $k_z$ directions by the shearing. For example, let $\mathbf{K}_1 = (0, 1, 0)$ and $\mathbf{K}_2 = (1, 1, 0)$, which are adjacent wavevectors at $T = 0$ as shown in the lower panels of Fig. \ref{fig2}. At $T$ = 2, Eqs. (\ref{eq_keff}) and (\ref{eq_bigk}) give $\mathbf{K}_1 = (-3, 1, 0)$ and $\mathbf{K}_2 = (-2, 1, 0)$ which are still contiguous. For this reason we compute the phase differences between Fourier modes with the same value of $k_y$ following Eqs. (\ref{eq_deltaphikx}) and (\ref{eq_deltaphikz}), and then obtain Eqs.(\ref{eq_sphikx})-(\ref{eq_sphi}).

\begin{figure}
  \includegraphics[width=\textwidth]{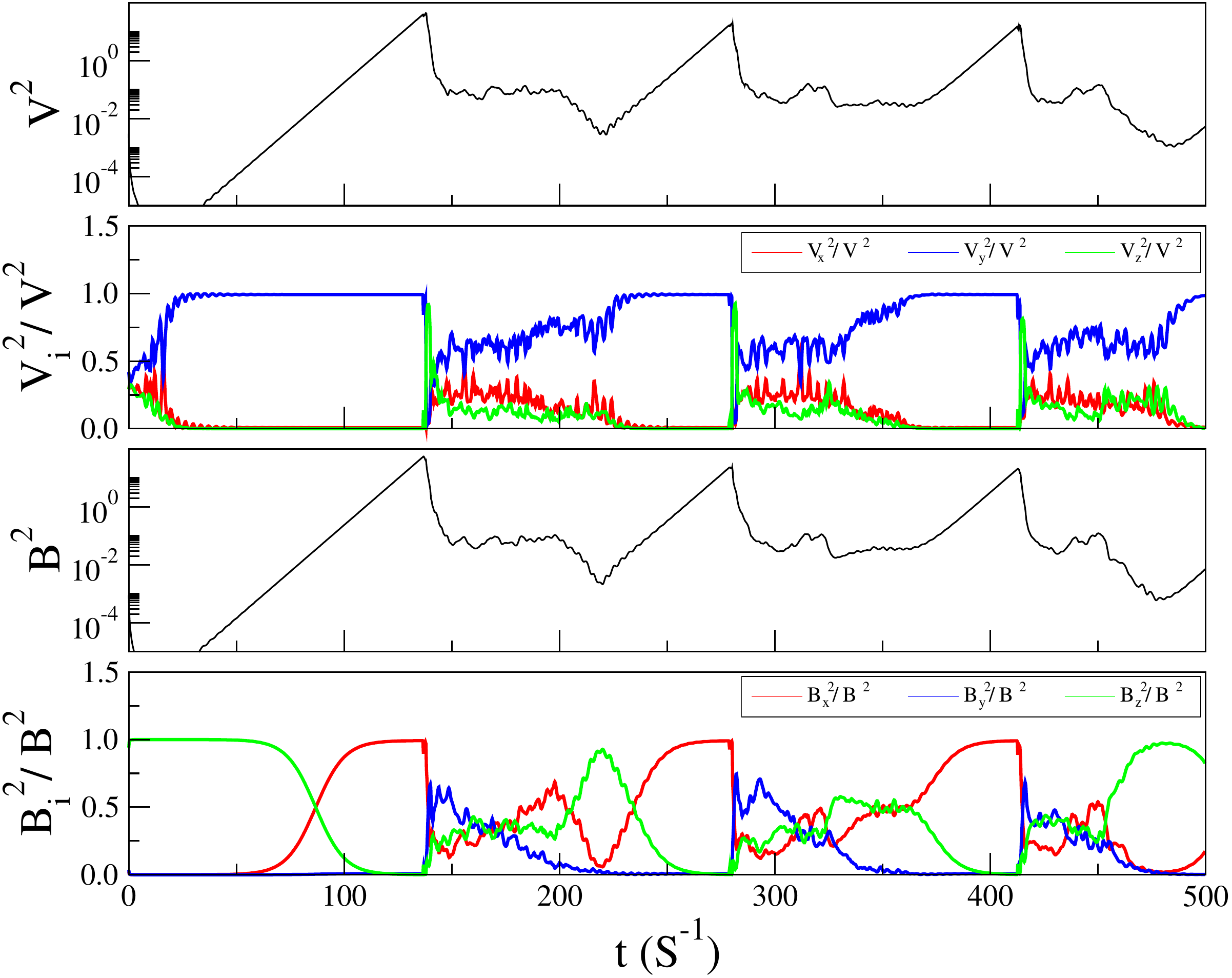}
  \caption{From top to bottom: time series of the total kinetic energy $V^2$, normalized kinetic energy of velocity components $V_i^2/V^2$, time series of the total magnetic energy $B^2$, and normalized energy of magnetic components $B_i^2/B^2$, $i = x, y, z$ ($\beta$ = 30).}
  \label{fig_ratios}
\end{figure}

\section{Numerical results} \label{sec_numerical}

We apply the Fourier power and phase spectral entropies to the $V_y$ component of the velocity field and the $B_x$ component of the magnetic field for $\beta$ = 30 and $\beta$ = 100. We choose $V_y$ and $B_x$ because they concentrate most of the total kinetic and magnetic energy, respectively, during the emergence of the channel flows at $\beta = 30$. This is illustrated in Fig. \ref{fig_ratios}. The upper panel shows the time series of the total kinetic energy

\begin{displaymath}
  V^2 = \frac{1}{L_x L_y L_z} \int_0^{L_z} \int_0^{L_y} \int_0^{L_x} |\mathbf{V}|^2 dx dy dz,
\end{displaymath}
\noindent and depicts a typical time series with large intermittent peaks due to the recurrent emergence of two-channel flows. The second panel shows the time series of the normalized kinetic energy of each component of the velocity field

\begin{displaymath}
  \frac{V_i^2}{V^2} = \frac{1}{L_x L_y L_z V^2} \int_0^{L_z} \int_0^{L_y} \int_0^{L_x} |V_i|^2 dx dy dz, \qquad i = x, y, z.
\end{displaymath}

\noindent The quantity $V_i^2/V^2 \in$ [0, 1], where 0 means that the $V_i$ component does not contribute to the total kinetic energy, and 1 indicates that the total kinetic energy is concentrated in the $V_i$ component. From Fig. \ref{fig_ratios} it is evident that almost all the kinetic energy is concentrated in the $V_y$ component during the formation of the large intermittent peaks at $\beta = 30$. The third panel of Fig. \ref{fig_ratios} shows the time series of the total magnetic energy

\begin{displaymath}
  B^2 = \frac{1}{L_x L_y L_z} \int_0^{L_z} \int_0^{L_y} \int_0^{L_x} |\mathbf{B}|^2 dx dy dz,
\end{displaymath}

\noindent and the bottom panel of Fig. \ref{fig_ratios} shows the time series of the normalized energy of each component of the magnetic field

\begin{displaymath}
  \frac{B_i^2}{B^2} = \frac{1}{L_x L_y L_z B^2} \int_0^{L_z} \int_0^{L_y} \int_0^{L_x} |B_i|^2 dx dy dz, \qquad i = x, y, z.
\end{displaymath}
\noindent It follows that, at $\beta = 30$, the $B_x$ component is responsible for most of the total magnetic energy during the growth of the large intermittent peaks. Therefore, we select the $V_y$ and $B_x$ component for our analysis. Nevertheless, similar results can be obtained for other components of the velocity and magnetic fields.

\subsection{$\beta$ = 30}

\begin{figure}
  \includegraphics[height=0.6\textheight]{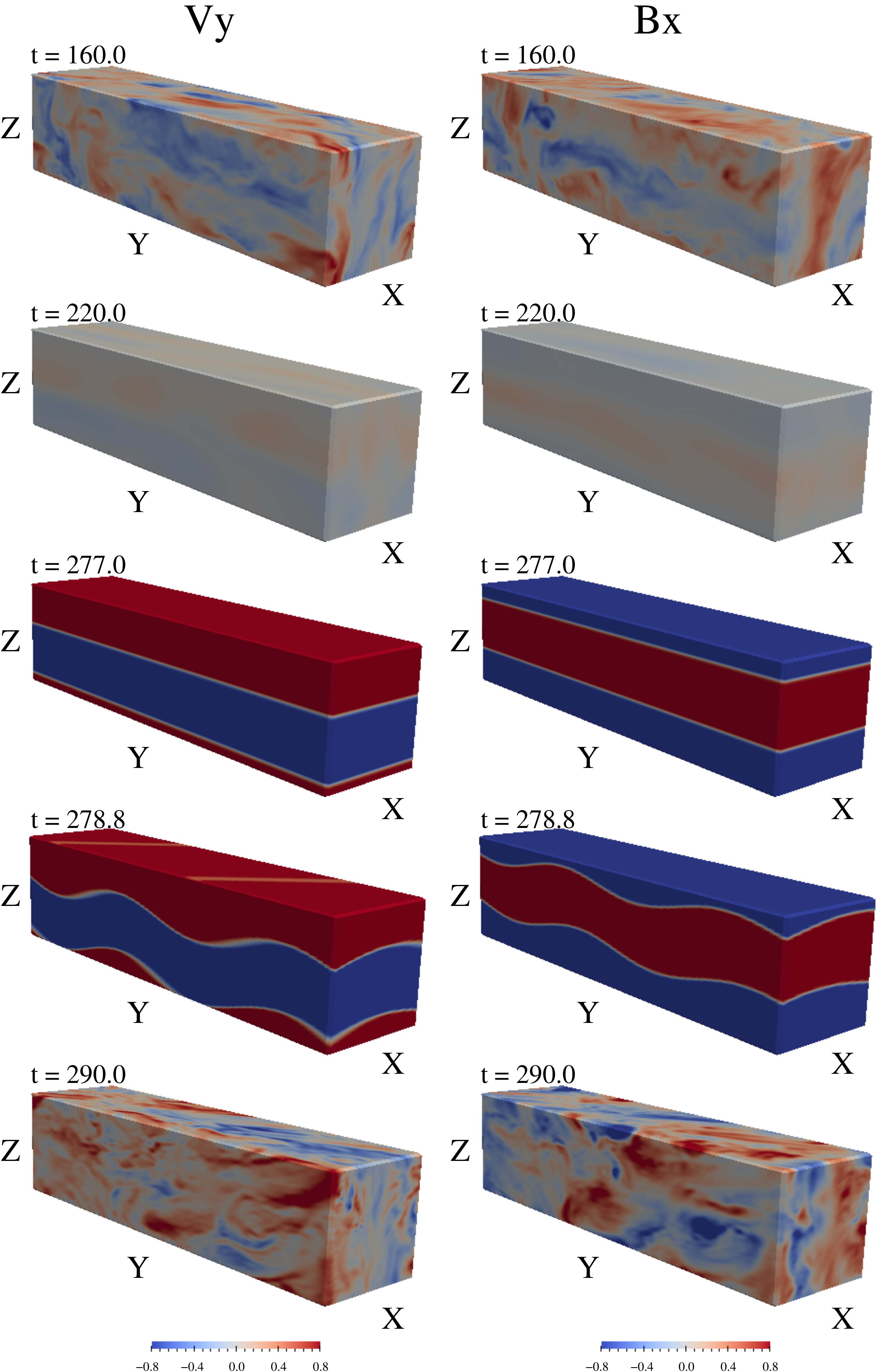}
  \caption{Spatiotemporal patterns of $V_y$ and $B_x$ at $t$ = 160 (turbulent state), $t$ = 220 (emergence of coherent structure), $t$ = 277 (large-amplitude coherent structure) $t$ = 278.8 (growing secondary instability) and $t$ = 290 (turbulent state) for $\beta = 30$.}
  \label{fig3}
\end{figure}

This regime is right above the threshold where the MRI takes place, and is characterized by the emergence of recurrent two-channel flows that can be easily recognized in the spatiotemporal patterns of the $V_y$ and $B_x$ components of the velocity and magnetic fields, respectively. At $t$ = 160 the patterns shown in the upper panels of Figure \ref{fig3} reflect a turbulent state. At $t$ = 220 the flow self-organizes forming a large-scale coherent pattern whose energy grows with time. At $t$ = 277 the two-channel flow reaches its maximum energy before it breaks at $t$ = 278.8, generating turbulent patterns as displayed in the bottom panels of Fig. \ref{fig3} for $t$ = 290. This figure shows that the velocity and magnetic fields display recurrent generation and breaking of a large-scale coherent structure (i.e., the two-channel flow).

\begin{figure}
  \includegraphics[width=\textwidth]{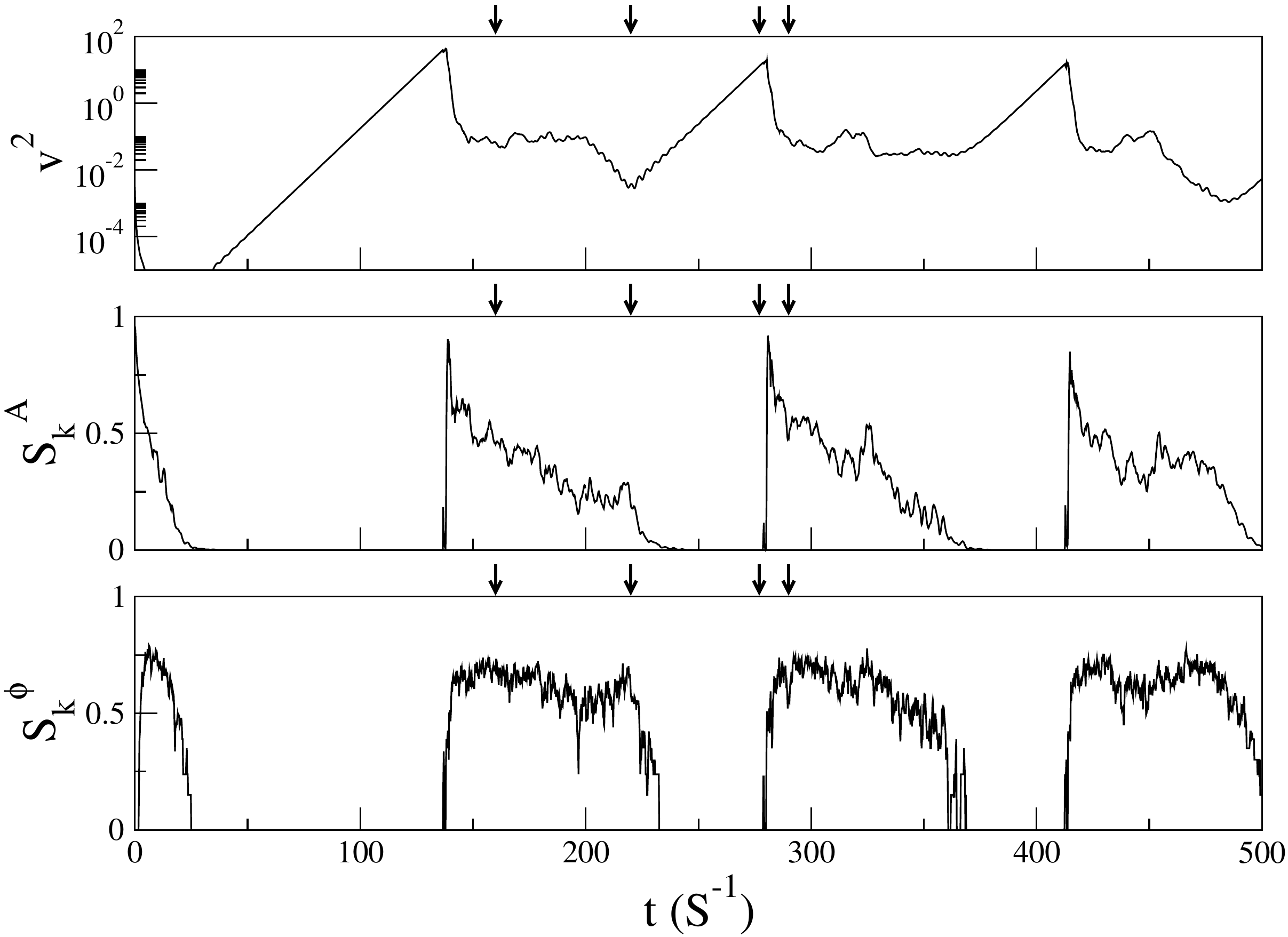}
  \caption{Time series of the kinetic energy (upper panel), the Fourier power spectral entropy (middle panel) and the Fourier phase spectral entropy (bottom panel) for $\beta = 30$.}
  \label{fig4}
\end{figure}

The upper panel of Fig. \ref{fig4} shows the time series of the total kinetic energy. Arrows indicate the selected values of $t$ shown in Fig. \ref{fig3}. The middle panel of Fig. \ref{fig4} shows the time series of the Fourier power spectral entropy $S_k^A$, and the bottom panel shows the time series of the Fourier phase spectral entropy $S_k^\phi$. Throughout our analysis, the power and phase spectral entropies have been normalized to their maximum values pointed out in Section \ref{sec_entropies}, thus $S_k^A, S_k^\phi \in [0, 1]$. Clearly, the peaks of the kinetic energy coincide with null values of the Fourier power and phase spectral entropies, which is expected since for such low $\beta$ only the largest wavelength mode in the $z$ direction is unstable \citep{lesur_longaretti:2007}.

\begin{figure}
  \includegraphics[width=\textwidth]{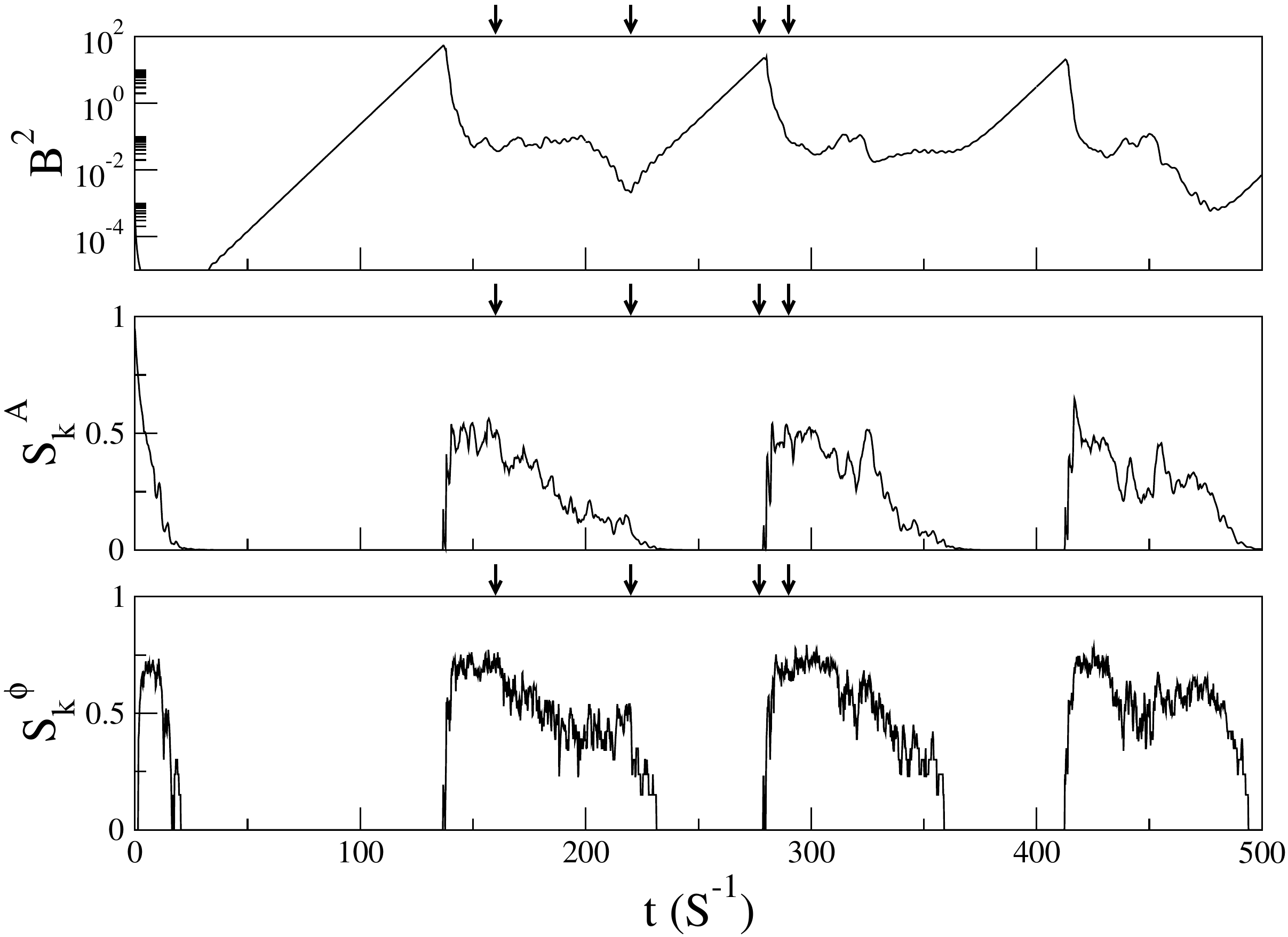}
  \caption{Time series of the magnetic energy (upper panel), the Fourier power spectral entropy (middle panel) and the Fourier phase spectral entropy (bottom panel) for $\beta = 30$.}
  \label{fig6}
\end{figure}

The time series of the total magnetic energy is displayed in the upper panel of Fig. \ref{fig6}. The middle panel shows the time series of the Fourier power spectral entropy $S_k^A$. Peaks in the time series of the magnetic energy are associated with null values of $S_k^A$. The bottom panel shows the time series of the Fourier phase spectral entropy $S_k^\phi$, which also displays null values associated with the peaks in the time series of the magnetic energy.

The sinusoidal shape of the coherent structure shown in Fig. \ref{fig3} at $t$ = 277 can be elucidated by computing the averages of the $V_y$ and $B_x$ fields in the $x$ and $y$ directions
  \begin{eqnarray*}
  \left< V_y \right>_{xy} (z, t) & = & \frac{1}{N_x N_y} \int_0^{L_y} \int_0^{L_x} V_y(x, y, z, t) dx dy \\
  \left< B_x \right>_{xy} (z, t) & = & \frac{1}{N_x N_y} \int_0^{L_y} \int_0^{L_x} B_x(x, y, z, t) dx dy
  \end{eqnarray*}

\begin{figure}
  \includegraphics[height=0.6\textheight]{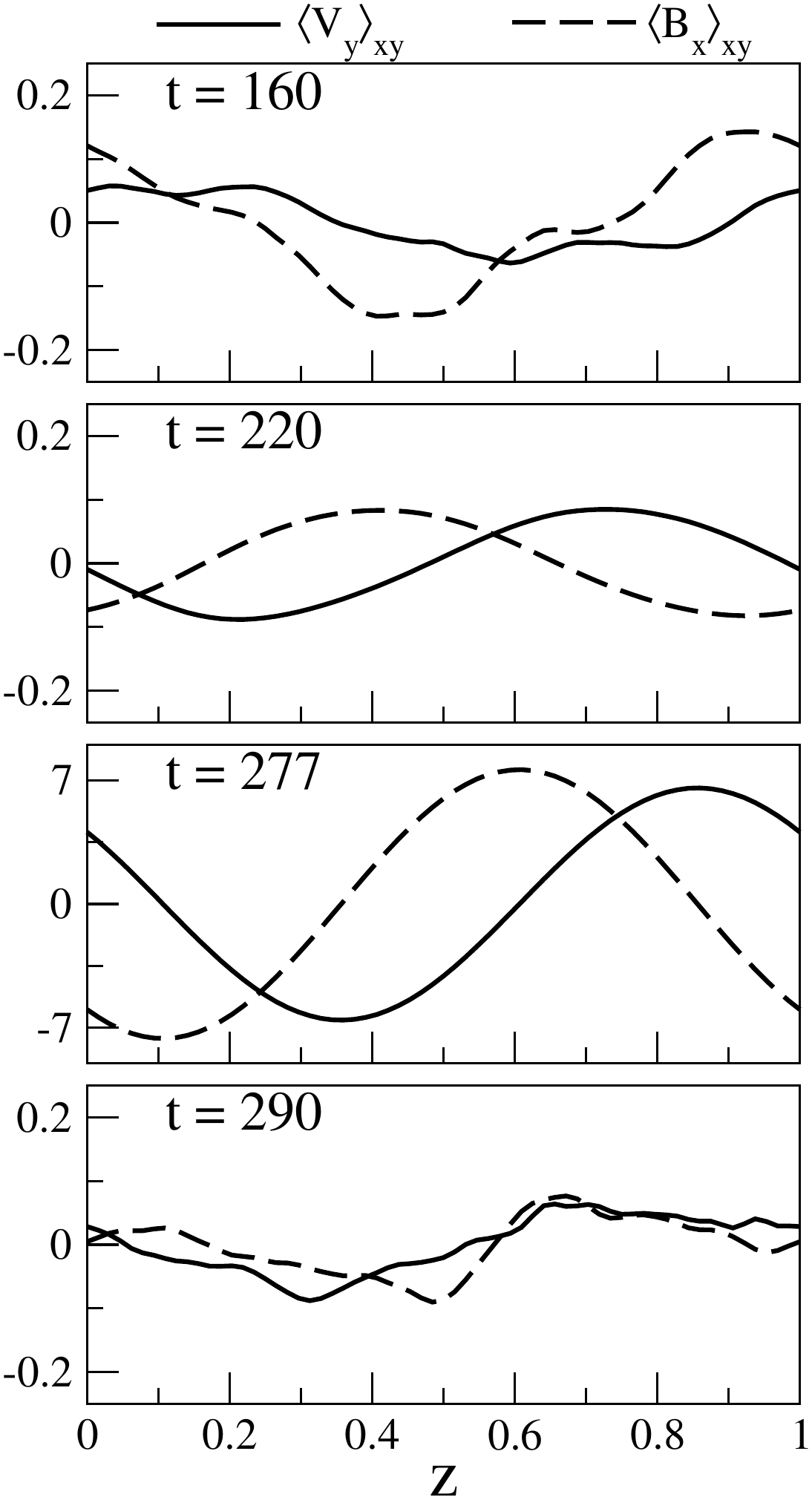}
  \caption{$\left< V_y \right>_{xy}$ (continuous line) and $\left< B_x \right>_{xy}$ (dashed line) as a function of $z$ at $t$ = 160, $t$ = 220, $t$ = 277 and $t$ = 290 for $\beta = 30$. Note the different scaling of the vertical axis for $t$ = 277.}
  \label{fig_spaceavrgs}
\end{figure}

\noindent Fig. \ref{fig_spaceavrgs} shows $\left< V_y \right>_{xy}$ and $\left< B_x \right>_{xy}$ as a function of $z$ for $t$ = 160, 220, 277 and 290. At $t$ = 160 (upper panel) the horizontal averages of $V_y$ and $B_x$ display irregular fluctuations owing to the turbulent state of the flow. At $t$ = 220 both curves exhibit sinusoidal shapes due to the formation of the two-channel flow, which grows until $t$ = 277. Note the phase shift between the velocity and magnetic fields. The average of $V_y$ and $B_x$ at $t$ = 278.8 is not shown because the undulations leading to the breaking of the two-channel flow are smoothed after the spatial averaging, and the resulting profiles of $\left< V_y \right>_{xy}$ and $\left< B_x \right>_{xy}$ are similar to those of $t$ = 277. After the destruction of the large-scale coherent structure the horizontal averages of $V_y$ and $B_x$ display irregular shapes seen at $t$ = 290.

Two-channel flows are exact solutions to the incompressible MHD equations \citep{goodman_xu:1994}, and can be represented by
  \begin{eqnarray*}
    \mathbf{V}^{ch}(z, t) & = & A_0 e^{\sigma t} \sin(k_z z) \left[\cos(\gamma) \mathbf{\hat{x}} + \sin(\gamma) \mathbf{\hat{y}} \right], \\
    \mathbf{B}^{ch}(z, t) & = & A_0 e^{\sigma t} \cos(k_z z) \left[\cos(\vartheta) \mathbf{\hat{x}} + \sin(\vartheta) \mathbf{\hat{y}} \right],
  \end{eqnarray*}

\noindent where $A_0$ is the amplitude at $t = 0$, $\sigma$ is the growth rate of the two-channel flow, $\gamma$ and $\vartheta$ represent the angle of the velocity and magnetic field vectors, respectively, relative to the $x$ axis in the $(x, y)$ plane. When $P_m$ = 1, the magnetic field is orthogonal to the velocity field \citep{pessah_chan:2008} thus $\vartheta = \gamma - \frac{\pi}{2}$, and

  \begin{eqnarray}
    \mathbf{V}^{ch}(z, t) & = & A_0 e^{\sigma t} \sin(k_z z) \left[\cos(\gamma) \mathbf{\hat{x}} + \sin (\gamma) \mathbf{\hat{y}} \right], \label{eq_vch1} \\
    \mathbf{B}^{ch}(z, t) & = & A_0 e^{\sigma t} \cos(k_z z) \left[\sin(\gamma) \mathbf{\hat{x}} - \cos (\gamma) \mathbf{\hat{y}} \right]. \label{eq_vch2}
  \end{eqnarray}

\noindent The orientation angle $\gamma$ can be computed as follows \citep{goodman_xu:1994}

  \begin{equation} \label{eq_atan_angle}
    \gamma = \tan^{-1} \left( \frac{ V^{ch}_y }{ V^{ch}_x} \right) = \tan^{-1} \left( \frac{ -B^{ch}_x }{ B^{ch}_y} \right),
  \end{equation}

\noindent where $V^{ch}_i$ and $B^{ch}_i$, $i = x, y$, represent the field component of the channel flow in the $i$-th direction. The growth rate $\sigma$ can be obtained by superposing (\ref{eq_vch1}) and (\ref{eq_vch2}) on the background shear velocity field and the vertical magnetic field, respectively

  \begin{eqnarray}
    \mathbf{u} & = & S x \mathbf{\hat{y}} + \mathbf{V}^{ch} \label{eq_chplusback1}, \\
    \mathbf{B} & = & B_0 \mathbf{\hat{z}} + \mathbf{B}^{ch} \label{eq_chplusback2}.
  \end{eqnarray}

\begin{figure}
  \includegraphics[height=0.9\textheight]{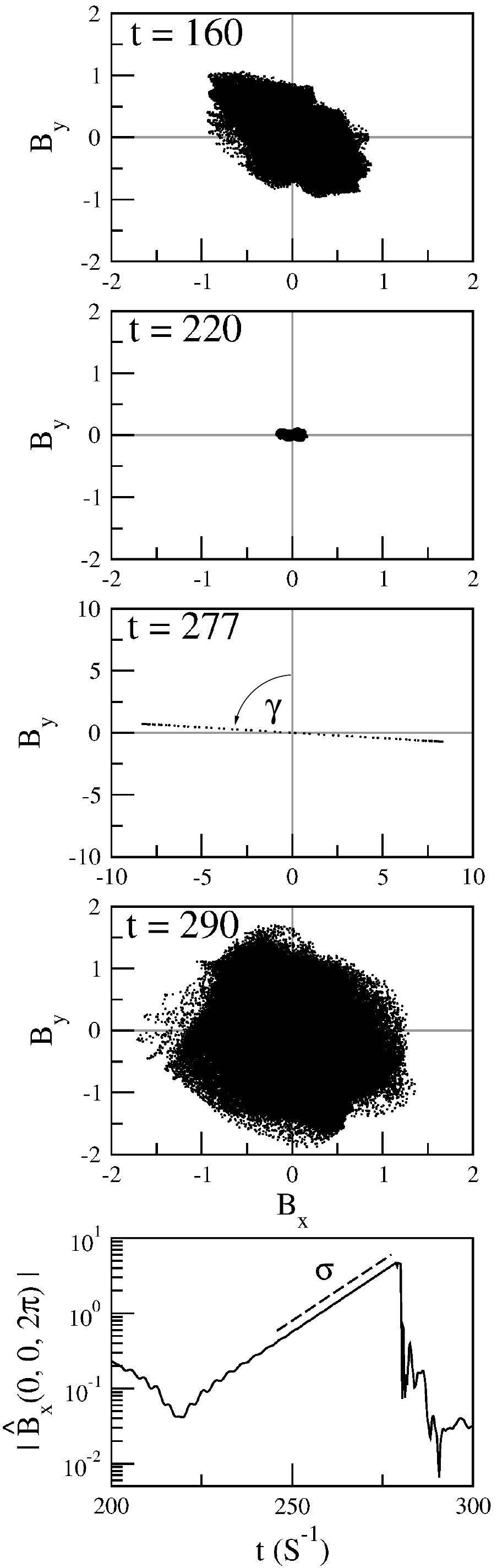}
  \caption{Scatter plot of the horizontal components ($B_x$, $B_y$) of the magnetic field for $\beta = 30$ at $t$ = 160, 220, 277 and 290. For $t$ = 277, $\gamma$ = 1.48. Note the different scaling for $t$ = 277. The lower panel shows the time series of the amplitude of $\hat{B}_z(0, 0, 2\pi)$, and the dashed line represents a least-squares fit with slope $\sigma$ = 0.074.}
  \label{figanglebeta30}
\end{figure}

\noindent Substituting Eqs. (\ref{eq_chplusback1}) and (\ref{eq_chplusback2}) into Eqs. (\ref{eq_mhd1}) and (\ref{eq_mhd2}), and neglecting the kinetic viscosity and the magnetic resistivity one obtains \citep{goodman_xu:1994}

  \begin{equation} \label{eq_growthrate_ch}
    \sigma = \frac{S}{2} \sin(2 \gamma).
  \end{equation}

\noindent The details are given in Appendix \ref{appendix_growthrate}. The growth rate $\sigma$ is maximum when $\gamma = \pi/4$, which by Eq. (\ref{eq_atan_angle}) implies that $V^{ch}_x = V^{ch}_y$ and $B^{ch}_{x} = -B^{ch}_{y}$. The growth rate decreases when $\gamma \rightarrow \pi/2$, which occurs when $V^{ch}_x \ll V^{ch}_y$ and $B^{ch}_{y} \ll B^{ch}_x$ in Eq. (\ref{eq_atan_angle}).

The analytical value of $\sigma$ can be compared with the growth rate of the large-scale coherent structure observed in the numerical simulation. The angle $\gamma$ in Eq. (\ref{eq_growthrate_ch}) can be obtained as follows. Recall that the spatial domain is discretized using a grid with resolution $64 \times 128 \times 64$, which gives 524288 datapoints for each component of the velocity and magnetic fields. By plotting the $B_y$ component of each datapoint as a function of its $B_x$ value and computing the slope of the resulting figure one can estimate the angle $\gamma$ (Eq. (\ref{eq_atan_angle})). This is illustrated in Fig. \ref{figanglebeta30} for $t$ = 160, 220, 277 and 290. At $t$ = 160, the points are scattered due to the turbulent patterns of the flow. At $t$ = 220 the region occupied by the ($B_x$, $B_y$) points is smaller, reflecting the initial formation of the coherent structure from small-amplitude fluctuations. At $t$ = 277 all points lie in a straight line, and most of them are located at the end points of the line, which is expected from a pure channel flow \citep{bodo_etal:2008}. From this panel we estimate the angle $\gamma$ using Eq. (\ref{eq_atan_angle}), which gives $\gamma = 1.48$ radians. Substituting into Eq. (\ref{eq_growthrate_ch}) we obtain the analytical value $\sigma = 0.086$. At $t$ = 290 the ($B_x$, $B_y$) points become scattered again, due to the breaking of the coherent structure and the flow returning to a turbulent state.

The growth rate $\sigma$ can also be estimated directly from the time series of the amplitude of the Fourier mode that corresponds to the channel flow \citep{sano:2007}. For $\beta = 30$, the channel flow is represented by a single mode with wavevector $\mathbf{k} = (k_x, k_y, k_z) = (0, 0, 2\pi)$ corresponding to $(n_x, n_y, n_z) = (0, 0, 1)$ (i.e., a Fourier mode in the vertical direction with the smallest $k_z$ wavenumber). The lower panel of Fig. \ref{figanglebeta30} shows the time series of the amplitude of $\hat{B}_x(0, 0, 2\pi)$. A least-squares fit of the time series from $t$ = 245 until $t$ = 277 with a function of the form $f(t) = A_0 \exp(\sigma t)$ gives $\sigma = 0.074$, which is close to the analytical result given by Eq. (\ref{eq_growthrate_ch}).


\begin{figure}
  \includegraphics[height=0.6\textheight]{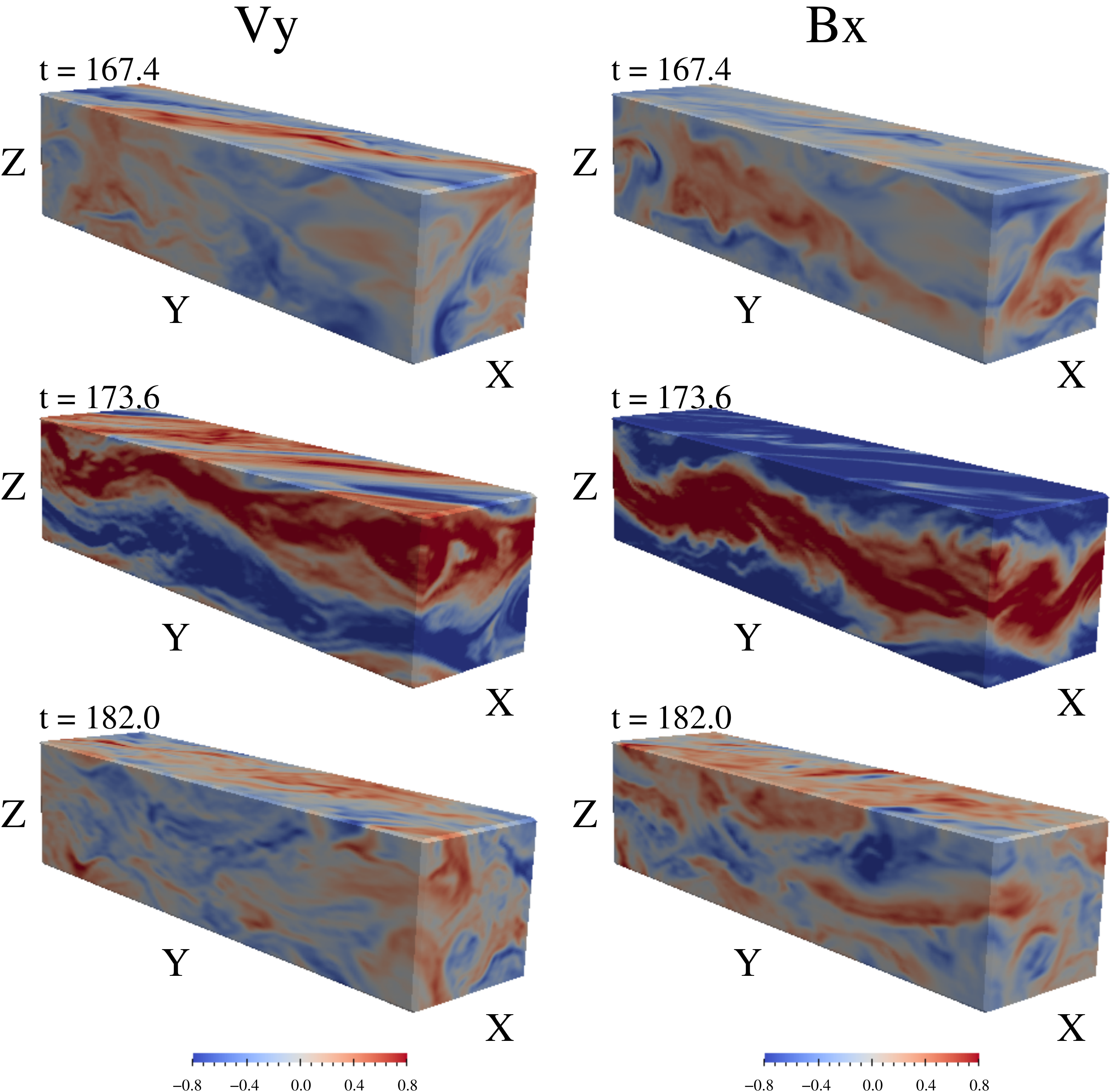}
  \caption{Spatiotemporal patterns of $V_y$ and $B_x$ at $t$ = 167.4, $t$ = 173.6 and $t$ = 182.0 for $\beta = 100$.}
  \label{fig7}
\end{figure}

\subsection{$\beta$ = 100}

In this regime turbulent patterns prevail, and there is sporadic emergence of coherent structures embedded in the turbulence. This is illustrated in Figure \ref{fig7} using the $V_y$ and $B_x$ components of the velocity and magnetic fields, respectively. The top panels show a typical turbulent pattern at $t$ = 167.4. The middle panels display a large-scale coherent structure emerging at $t$ = 173.6 from the turbulent pattern. At $t$ = 182 (bottom panel), the coherent structure has broken into smaller eddies and the patterns return to a turbulent state. Note that the coherent structure cannot be regarded as a pure channel flow, since several Fourier modes are actively contributing to the turbulent fluctuations that coexist with the coherent structure. Nevertheless, it keeps the basic shape of a two-channel flow. Its destruction can no longer be solely attributed to parasitic modes, but low-level turbulence plays a key role \citep{latter_etal:2009}. Perhaps, the best way to describe the formation of this structure is by measuring the degree of synchronization between modes, as done below.

\begin{figure}
  \includegraphics[width=\textwidth]{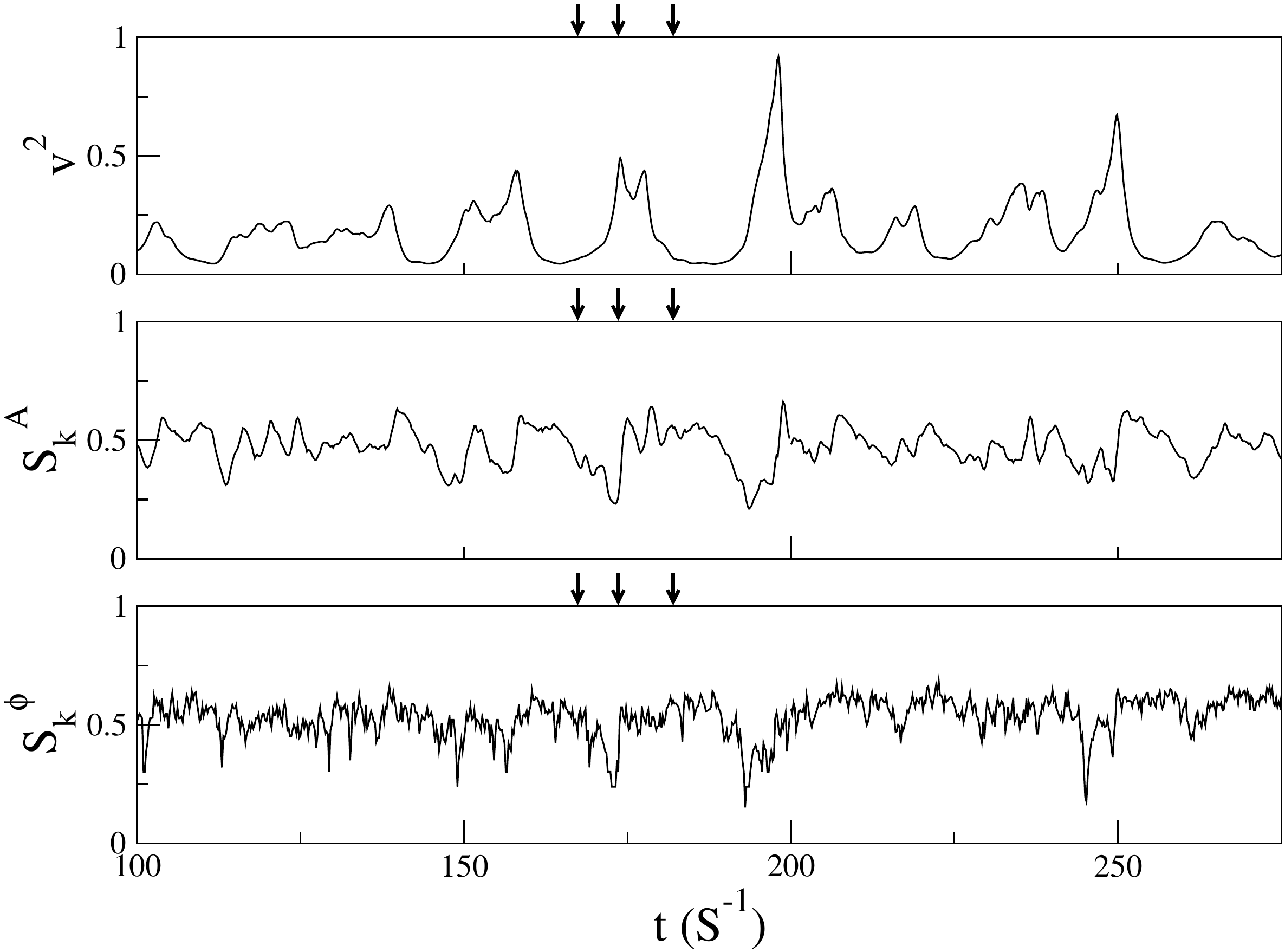}
  \caption{Time series of the kinetic energy (upper panel), the Fourier power spectral entropy (middle panel) and the Fourier phase spectral entropy (bottom panel) for $\beta = 100$. Arrows indicate the selected values of $t$ of the spatiotemporal patterns shown in Fig. \ref{fig7}.}
  \label{fig8}
\end{figure}

The time series of the kinetic energy is displayed in the upper panel of Fig. \ref{fig8}, with the values of $t$ displayed in Fig. \ref{fig7} being indicated by arrows. Note that the appearance of the coherent structure is followed by an increase in the kinetic energy. The same phenomenon is observed throughout the time series and had previously been reported by \citet{he_chian:2003} in the context of on-off collective imperfect phase synchronization in a one-dimensional wave equation. The time series of the Fourier power spectral entropy $S_k^A$ is shown in the middle panel. Peaks in the time series of the kinetic energy coincide with low values of $S_k^A$, indicating a decrease of spatial disorder during the emergence of the coherent structures. The bottom panel shows the time series of the Fourier phase spectral entropy $S_k^\phi$, which displays higher variability than $S_k^A$. However, the strong drops in $S_k^\phi$ are correlated with corresponding drops in $S_k^A$, in particular during the interval indicated by the arrows. From this figure we can conclude that the embedded coherent structures in this regime are characterized by high degrees of amplitude-phase synchronization. Thus, the on-off intermittency can be described as follows. In ``on'' stages, the Fourier modes associated with different spatial scales adjust themselves to collective amplitude-phase synchronization, inducing bursts in the wave energy, since the mode energies are effectively superimposed. In ``off'' stages, the synchronization is weaker and the energy becomes lower.

\begin{figure}
  \includegraphics[width=\textwidth]{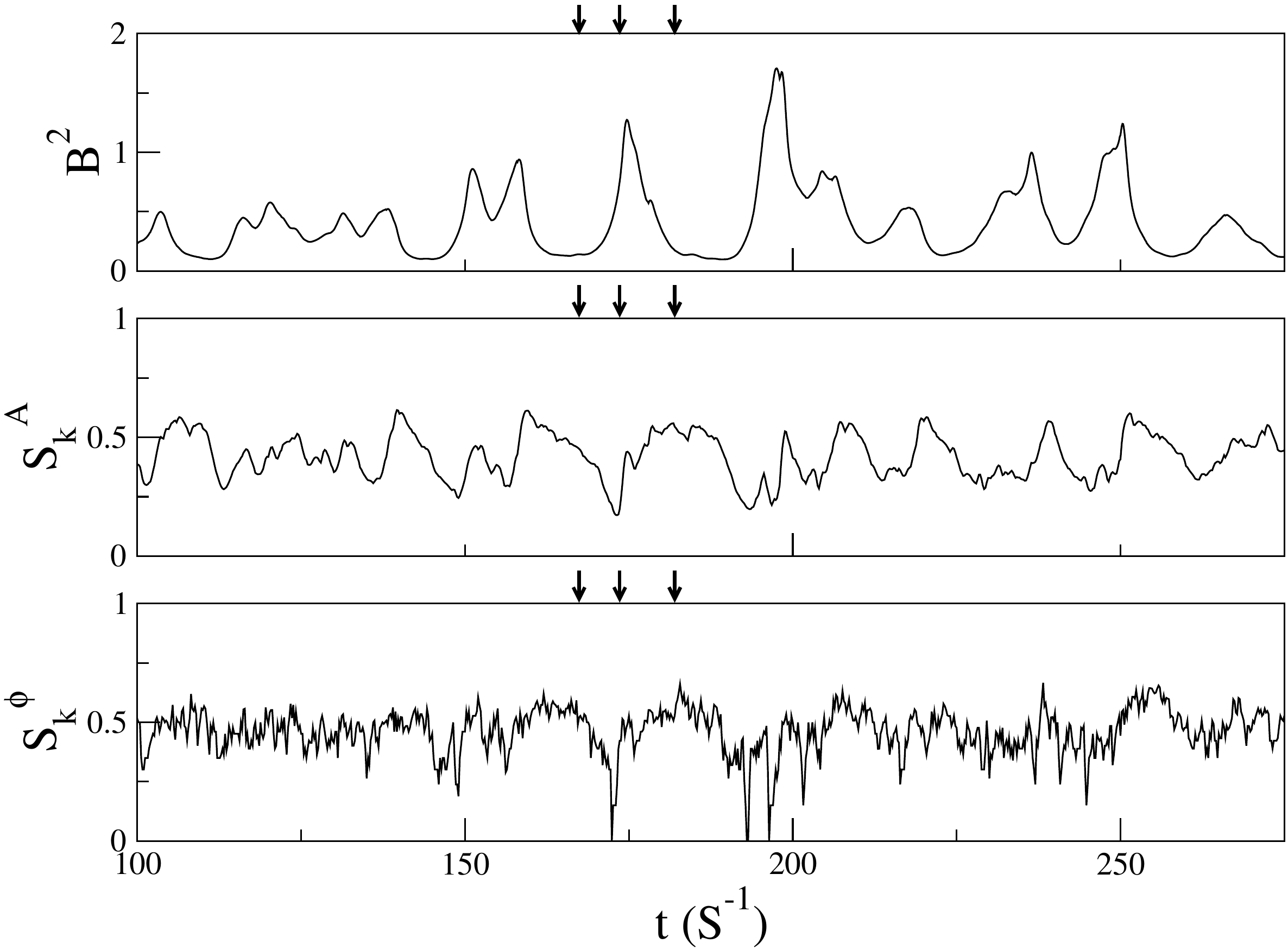}
  \caption{Time series of the magnetic energy (upper panel), the Fourier power spectral entropy (middle panel) and the Fourier phase spectral entropy (bottom panel) for $\beta = 100$. Arrows indicate the selected values of $t$ of the spatiotemporal patterns shown in Fig. \ref{fig7}.}
  \label{fig10}
\end{figure}

We repeat the same analysis to the magnetic field. Fig. \ref{fig10} shows the time series of the magnetic energy, the Fourier power spectral entropy $S_k^A$ and the Fourier phase spectral entropy $S_k^\phi$. Arrows indicate the selected values of $t$ shown in Fig. \ref{fig7}. These results lead us to the same conclusion as the velocity field, that the coherent structures in the magnetic field are the result of amplitude-phase synchronization among scales during the ``on'' stages of the on-off intermittency.

\section{Discussion} \label{sec_discussion}


The emergence of the two-channel flow in Keplerian shear flows with a vertical background magnetic field has attracted much attention since it was observed in the numerical simulations of \citet{hawley_balbus:1992} and later recognized as exact nonlinear solutions of the MHD equations in the theoretical work of \citet{goodman_xu:1994}. Using a semi-analytical analysis, \citet{goodman_xu:1994} also demonstrated that two-channel flows are subject to secondary, ``parasitic'' instabilities with growth rates proportional to the MRI. The mechanism through which the two-channel flow grows and is destroyed has been discussed by several authors \citep{goodman_xu:1994, latter_etal:2009, pessah_goodman:2009, longaretti_lesur:2010}. According to one scenario that is consistent with our simulations at $\beta = 30$, the channel mode exponentially grows from random, small-scale white noise. The amplitude of the channel mode continues to grow during several shear time units because the ``narrow box'' geometry of the spatial domain adopted in this paper can suppress the fastest-growing parasitic instabilities, which have long wavelengths in the horizontal direction \citep{bodo_etal:2008}. At some amplitude, the growth of the channel mode is halted by a parasitic instability and is destroyed. Presumably, the parasitic modes themselves decay into small-scale turbulence, which may then produce the seed for the random fluctuations out of which the channel flow grows \citep{longaretti_lesur:2010}.

The recurrence of the patterns observed in our numerical simulations of Keplerian shear flows resembles the nonlinear behaviour of dynamical systems at the onset of spatiotemporal chaos. Numerical solutions of one-dimensional models of drift waves in plasmas display intermittent switching between two nonattracting coherent structures known as chaotic saddles \citep{rempel_chian:2007, rempel_etal:2007, rempel_etal:2009a, chian_etal:2010}. A similar behavior was reported in numerical simulations of nonlinear MHD dynamos in compressible flows \citep{rempel_etal:2009b}. We presume that, in the context of dynamical systems, the trajectory given by the numerical solution of the shearing-box equations with small-scale noise as initial conditions approaches a region of the phase space dominated by the channel mode, and then escapes toward a region of the phase space near a chaotic set that represents the turbulent state. This behavior can depend on the initial conditions. \citet{balbus_lessaffre:2008} and \citet{lessaffre_etal:2009} derived a low-dimensional model of MRI-driven turbulence and found that the phase space is dominated by three fixed points: a stable fixed point that represents a ``turbulent'' state, a stable fixed point that represents a ``quiet'' state, and an unstable fixed point that separates the basins of attraction of the ``turbulent'' and ``quiet'' states. They compared their reduced model with 3D simulations of shear flows, and found that their model can predict the final state of the shearing-box simulations. On-off intermittency is a common phenomenon of dynamical systems, and a detailed analysis of the simulations presented here employing a dynamical systems approach to understand the mechanism responsible for the recurrent patterns observed will be the topic of future work.

The Fourier phase spectral entropy defined in this paper is computed using phase differences between adjacent modes (Eqs. (\ref{eq_deltaphikx}) and (\ref{eq_deltaphikz})), which means that it is restricted to local interactions between scales. The transfer of energy in MHD turbulence can be non-local \citep{brandenburg:2001, lesur_longaretti:2011}. Therefore, it would be interesting to modify Eqs. (\ref{eq_deltaphikx}) and (\ref{eq_deltaphikz}) to take into account non-local interactions between Fourier modes. For example, in the particular case of MHD turbulence in Keplerian shear flows with a background magnetic field, the faster-growing parasitic instabilities have large horizontal wavelengths, which correspond to small values of $k_x$ and $k_y$, thus one can identify the parasitic modes in Fourier space and modify the definition of the Fourier phase entropy to compute the phase differences between the channel modes and the parasitic modes, directly quantifying their synchronization.


\section{Conclusion} \label{sec_conclusion}

In summary, in this paper we applied the Fourier power and phase spectral entropies to quantify the degree of amplitude-phase synchronization due to multiscale interactions in a three-dimensional model of magnetized Keplerian shear flows. We have defined carefully the Fourier power and phase spectral entropies for 3D fields, and the phase differences between neighbouring modes in the presence of shearing boundary conditions. Then, we focused on two different values of $\beta$. The first regime ($\beta$ = 30) is characterized by turbulent patterns that self-organize, leading to the recurrent formation of two-channel flows. We showed that two-channel flows are characterized by null values of the Fourier power and phase spectral entropies, which reflects the single-mode ordered state of their spatiotemporal patterns. In the second regime ($\beta$ = 100) the coherent structures are characterized by several Fourier modes and have more difficulty to stabilize once they emerge from the turbulent background. Their breaking mechanism is qualitatively different from the $\beta = 30$ case. Their formation is again related to low values of Fourier power and phase spectral entropies, but since there is more than a single mode involved in the dynamics of the coherent structures, this result indicates an increase of amplitude-phase synchronization of modes during the ``on'' stages of the on-off intermittency. Conversely, a loss of synchronization is responsible for the breaking of the structures and the beginning of the ``off'' stages.

The Fourier power spectral entropy and the Fourier phase spectral entropy, as defined in Eqs. (\ref{eq_ska}) and (\ref{eq_sphi}) respectively, can be useful to quantify the degree of amplitude-phase synchronization due to multiscale interactions. On-off intermittency is a phenomenon present in a wide range of complex systems, including numerical simulations of the solar dynamo \citep{sweet_etal:2001, brandenburg_spiegel:2008, spiegel:2009, rempel_etal:2009a, rempel_etal:2011}, spherical Couette dynamos \citep{raynaud_dormy:2013} and pulsar magnetospheres \citep{li_etal:2012}. Therefore, the techniques described in this work can be readily applied to the investigation of other turbulent systems relevant to space and astrophysical plasmas.

\section*{Acknowledgments}
The authors thank the reviewer for valuable comments. This work is supported by CNPq, FAPESP and DPP/UnB in Brazil.

\appendix

\section{The growth rate of the two-channel flow} \label{appendix_growthrate}

The growth rate of the two-channel flow can be obtained by proposing solutions of Eqs. (\ref{eq_mhd1})-(\ref{eq_mhd4}) of the form \citep{goodman_xu:1994,latter_etal:2009}

  \begin{eqnarray}
    \mathbf{V}^{ch} & = & A_0 e^{\sigma t} \sin(k_z z) \left[ \cos(\gamma) \mathbf{\hat{x}} + \sin(\gamma) \mathbf{\hat{y}} \right], \label{eq_appendix_vch}\\
    \mathbf{B}^{ch} & = & A_0 e^{\sigma t} \cos(k_z z) \left[ \sin(\gamma) \mathbf{\hat{x}} - \cos(\gamma) \mathbf{\hat{y}} \right], \label{eq_appendix_bch}
  \end{eqnarray}

\noindent where $A_0$ is the amplitude at $t$ = 0, $\sigma$ is the growth rate, and $\gamma$ is the orientation angle. Here we assume that $Re = Rm$, or equivalently, $Pm = 1$. By superimposing (\ref{eq_appendix_vch}) on the background shear flow

  \begin{equation} \label{eq_appendix_v}
    \mathbf{u} = \mathbf{u}_0 + \mathbf{V}^{ch} = -Sx\mathbf{\hat{y}} + \mathbf{V}^{ch},
  \end{equation}

\noindent and (\ref{eq_appendix_bch}) on the background vertical magnetic field
  \begin{equation} \label{eq_appendix_b}
    \mathbf{B} = B_0 \mathbf{\hat{z}} + \mathbf{B}^{ch},
  \end{equation}

\noindent and substituting (\ref{eq_appendix_v}) and (\ref{eq_appendix_b}) into the induction equation (\ref{eq_mhd2}) we obtain

  \begin{equation} \label{eq_appendix_induction_ch}
    \frac{ \partial \mathbf{B}^{ch}}{\partial t} = B_0 \frac{\partial \mathbf{V}^{ch}}{\partial z} - S B_x^{ch} \mathbf{\hat{y}},
  \end{equation}

\noindent where we have neglected the kinetic viscosity and magnetic resistivity. By inserting Eqs. (\ref{eq_appendix_vch}) and (\ref{eq_appendix_bch}) into (\ref{eq_appendix_induction_ch}) we have

  \begin{displaymath}
    \sigma A_0 e^{\sigma t} \cos(k_z z) \left[\sin(\gamma) \mathbf{\hat{x}} - \cos(\gamma) \mathbf{\hat{y}} \right] = B_0 k_z A_0 e^{\sigma t} \cos(k_z z) \left[ \cos(\gamma) \mathbf{\hat{x}} + \sin(\gamma) \mathbf{\hat{y}} \right] - S A_0 e^{\sigma t} \cos(k_z z) \sin(\gamma) \mathbf{\hat{y}}.
  \end{displaymath}

\noindent Separating into components

  \begin{eqnarray*}
     \sigma A_0 e^{\sigma t} \cos(k_z z) \sin(\gamma) & = & B_0 k_z A_0 e^{\sigma t} \cos(k_z z) \cos(\gamma),\\
    -\sigma A_0 e^{\sigma t} \cos(k_z z) \cos(\gamma) & = & B_0 k_z A_0 e^{\sigma t} \cos(k_z z) \sin(\gamma) - S A_0 e^{\sigma t} \cos(k_z z) \sin(\gamma),
  \end{eqnarray*}

\noindent and removing the term $A_0 e^{\sigma t} \cos(k_z z)$ we obtain

  \begin{eqnarray}
     \sigma \sin(\gamma) & = & B_0 k_z \cos(\gamma), \label{eq_appendix_compx}\\
    -\sigma \cos(\gamma) & = & B_0 k_z \sin(\gamma) - S \sin(\gamma). \label{eq_appendix_compy}
  \end{eqnarray}

\noindent Multiplying (\ref{eq_appendix_compx}) by $\sin(\gamma)$ and (\ref{eq_appendix_compy}) by $\cos(\gamma)$, and applying the trigonometric identity $2\sin(\gamma)\cos(\gamma) = \sin(2\gamma)$ we obtain

  \begin{displaymath}
    \sigma = \frac{S}{2} \sin(2\gamma).
  \end{displaymath}

\bsp

\label{lastpage}

\end{document}